%% file: root.tex
\documentclass[10pt,a4paper, titlepage]{article}

\usepackage[a4paper, total={6in, 8.5in}]{geometry}

\usepackage{amssymb}

\usepackage[figuresright]{rotating}

\usepackage[numbers]{natbib}
\usepackage{graphicx,caption,subcaption}

\usepackage[title]{appendix}
\usepackage{array}
\usepackage[table]{xcolor}
\usepackage{cuted}
\usepackage{algorithm}
\usepackage{algpseudocode}
\usepackage{amsmath,amsfonts}
\usepackage{mathtools}
\usepackage{amssymb}
\usepackage{enumitem}
\usepackage{epstopdf} 
\usepackage{tikz}
\usepackage{balance}
\usepackage{todonotes}
\usepackage{tabularx}
\usepackage{booktabs}
\usetikzlibrary{calc,arrows,matrix,positioning, fit, trees}

\newcommand{\lcol}[1]{\multicolumn{1}{|c}{#1}}    

\newcommand{\ws}{\textcolor{white}{I}} 
\newcommand{\nI}{\textcolor{black}{-I}} 

\providecommand{\keywords}[1]{\textbf{\textit{Keywords:}} #1}

\begin{document}





\title{Nonlinear Predictive Control on a Heterogeneous Computing Platform}

\author{Bulat Khusainov$^*$, Eric Kerrigan$^*{^\dagger}$, Andrea Suardi$^*$, George Constantinides$^*$  \\
\\
$^*$Department of Electrical \& Electronic Engineering\\ 
${^\dagger}$Department of Aeronautics\\
Imperial College London\\
London, SW7~2AZ, UK\\ 
\\
${^\dagger}$corresponding author: e.kerrigan@imperial.ac.uk
}

\maketitle


%

\begin{abstract}
\input{abstract.tex}
\end{abstract}





\input{introduction.tex}
\input{ocp_forumulation.tex}
\input{nmpc_algorithms.tex}
\input{implementation.tex}
\input{experimental_setup.tex}
\input{experimental_results.tex}
\input{conclusion.tex}
\input{acknowledgements.tex}

\begin{appendices}
\input{appendix_algorithm_details.tex}
\input{appendix_synthesis_results.tex}
\end{appendices}



\bibliography{root}







\end{document}

%% file: abstract.tex

We propose an implementation of an interior-point-based nonlinear predictive controller on a heterogeneous processor. The workload can be split between a general-purpose CPU and a field-programmable gate array to trade off the contradicting design objectives of control performance and computational resource usage. A new way of exploiting the structure of the KKT matrix yields significant memory savings. We report an 18x memory saving, compared to existing approaches, and a 36x speedup over a software implementation with an ARM Cortex-A9 processor. We also introduce a new release of Protoip, which abstracts low-level details of heterogeneous programming and allows processor-in-the-loop verification.

\keywords{Predictive control, Hardware-software co-design, Scheduling; FPGA; Optimization-based control}

%% file: introduction.tex
\section{Introduction}

Explicit performance optimization, systematic constraint handling and the inherent capability of dealing with nonlinearities are the main features that explain the success of Model Predictive Control (MPC) in recent decades~\citep{mayne_book}. In the MPC framework a dynamic optimization problem is solved at each sampling instant, which might restrict the application scope to systems with slow dynamics and/or render expensive implementations limited to high-performance computers. 

Conventionally these challenges were addressed on the algorithmic and software levels by developing new optimization problem formulations and generating hardware-efficient code~\citep{acado_reference}. However, in addition to improvements on the software side, recent developments in reconfigurable computing allowed acceleration of predictive control algorithms on Field-Programmable Gate Arrays (FPGAs), which resulted in low-cost and resource-efficient realizations of custom quadratic programming (QP) solvers for MPC~\citep{jerezetalIET:2012,hartley_fpga}. Extending a hardware acceleration approach from linear to nonlinear model predictive control (NMPC), which can be considered as the next logic step, requires mapping numerical integration algorithms on hardware, which is not a trivial task, since dynamic models that describe the physical world are problem-dependent. As a result, instead of using systematic dynamic optimization, existing FPGA implementations of NMPC either rely  on stochastic optimization~\citep{fpga_pso_nmpc} or approximate the offline solution with machine learning techniques~\citep{fpga_ann_nmpc}. These approaches cannot guarantee scalability nor closed-loop performance. 
  
Accelerating deterministic algorithms on hardware might be achieved by employing heterogeneous computing platforms that involve both a general-purpose processor with a fixed architecture and FPGA logic. For example, \citep{peyrl_hybrid_nmpc} present a heterogeneous implementation of a multiple-shooting based NMPC algorithm. The authors propose implementing integration in software while accelerating a fast gradient-based quadratic programming solver on an FPGA. The reported speedup of the heterogeneous implementation over a software realization is 1.6x and further improvement is limited, since integration and optimization algorithms have comparable computational complexity. This is a consequence of Amdahl's law~\citep{amdahls_law}, which states that an algorithm's speedup is limited by the part of the workload that cannot benefit from acceleration.

We present a new heterogeneous implementation of nonlinear interior-point algorithm for predictive control, that was first introduced in~\cite{bulat_heterogeneous_nmpc}. The main features of the proposed implementation are:
\begin{itemize}
\item{A method for scheduling sparse matrix-vector multiplication within an iterative linear system solvers to enable significant improvements in terms of computation time vs resource usage. For the example considered, an 18x memory saving compared to existing approaches and a 36x speedup over a software implementation are reported.}
\item{Flexible splitting of the algorithmic workload between software and hardware for trading off the computational resource usage against performance.}
\end{itemize}

In addition to the initial results presented in~\cite{bulat_heterogeneous_nmpc}, this paper presents the following extensions:
\begin{itemize}
	\item The whole family of implicit and explicit Runge-Kutta methods are supported for ordinary differential equations integration. In contrast, the initial implementation was limited to Euler method only.
	\item The optimal control objective is generalized from a quadratic function to nonlinear least squares.
	\item The proposed controller is experimentally validated in the loop with a gantry crane high-fidelity Simscape~\cite{matlab_bib} model. In~\cite{bulat_heterogeneous_nmpc} the controller was only validated in the loop with a nominal model.
\end{itemize}

Another contribution of the paper is a new release of the Protoip software tool~\citep{protoip}. Protoip allows quick prototyping and processor-in-the-loop verification of optimization algorithms on a Xilinx Zynq system-on-a-chip (SoC), which contains an ARM processor and FPGA fabric. In contrast to the previous releases, which were  focused on pure FPGA implementations, the new version of Protoip allows the incorporation of  both an ARM processor and FPGA. Protoip can be used both for quick testing of the proposed implementation from the MATLAB environment  and for design and verification of other heterogeneous implementations.

The remainder of the paper is organised as follows: Section~2 describes the considered optimal control problem formulation; existing NMPC algorithms, with a focus on suitability for hardware implementation, are reviewed in Section~3; in Section~4 a heterogeneous computer-based implementation of NMPC is presented, followed by experimental setup description (Section~5) and experimental results (Section~6). Note, that Protoip is a part of the experimental setup and hence presented in the corresponding section. Section~7 concludes the paper.

%% file: ocp_forumulation.tex
\section{Optimal Control Problem formulation}

We consider nonlinear time-invariant systems that can be described as an ordinary differential equation (ODE) of the form
\begin{equation}
	\dot{x}(t) = f_{c}(x(t), u(t))  ,
\end{equation}
\noindent 
where $f_c: \mathbb{R}^n \times \mathbb{R}^m \rightarrow \mathbb{R}^n$.
We consider the  nonlinear optimal control problem (OCP) with initial state~$\hat{x}$ and prediction horizon $T$:
\begin{subequations} 
\label{cont_nonlin_ocp}
\begin{equation} 
  \begin{aligned}
 \underset{x, u, s}{\textrm{min}} \quad \int_0^T  ||h(x(t),u(t),s(t))||_2^2  dt + ||h_T(x(T),s_T)||_2^2
 \end{aligned}
\end{equation}
	\begin{align} 
	\textrm{subject to:} &  \quad x(0) = \hat{x},   \\
		 & \quad \dot{x}(t) = f_{c}(x(t), u(t)), && \forall t \in \lbrack 0,T \rbrack    \label{cont_nonlin_dynamics}  \\
		 & \quad q(x(t),u(t),s(t)) = 0, && \forall t \in \lbrack 0,T \rbrack   \label{slack_eq} \\
		 & \quad q_T(x(T),s_T) = 0,  \label{term_slack_eq}  \\
		 & \quad x_l \leq  x(t) \leq x_u, && \forall t \in \lbrack 0,T \rbrack \\
		 & \quad u_l \leq  u(t) \leq u_u, && \forall t \in \lbrack 0,T \rbrack \\
		 & \quad s_l \leq  s(t) \leq s_u, && \forall t \in \lbrack 0,T \rbrack \\
		 & \quad x_{Tl} \leq  x(T) \leq x_{Tu}, \\
		 & \quad s_{Tl} \leq  s_T \leq s_{Tu}, \\
		  \label{eq:terminal}
	\end{align}
\end{subequations}
\noindent
where $h: \mathbb{R}^n \times \mathbb{R}^m \times \mathbb{R}^{n_{S}} \rightarrow \mathbb{R}^{n_{h}}$ and $h_T: \mathbb{R}^n \times \mathbb{R}^{n_{ST}} \rightarrow \mathbb{R}^{n_{hT}}$. Slack trajectory $s$ and slack variable $s_T$ are introduced alongside with equations~\eqref{slack_eq} and~\eqref{term_slack_eq}, which is a common technique that allows for the handling of general nonlinear inequality constraints~\cite{ipopt}. The presented formulation can be generalized for time-varying reference tracking, which, as will be shown later, requires only changing the software part of the algorithm.

%% file: nmpc_algorithms.tex
\section{Nonlinear predictive control algorithms}

Direct solution of the continuous-time optimal control problem~\eqref{cont_nonlin_ocp} involves two main stages: integration, i.e.\ solving the ordinary differential equation (ODE), and optimization. 
Implementing integration on an FPGA is not desirable because of the following reasons:
\begin{itemize}
\item The ODE~\eqref{cont_nonlin_dynamics} may involve mathematical expressions (e.g.\ sine and square root) that in contrast to standard addition and multiplication operations, require significant amounts of computational resources (Figure~\ref{fig:fpga_resource}) and might be unsuitable for pipelining.
\begin{figure}[tb]
	\centering
	\includegraphics[width=0.6\textwidth]{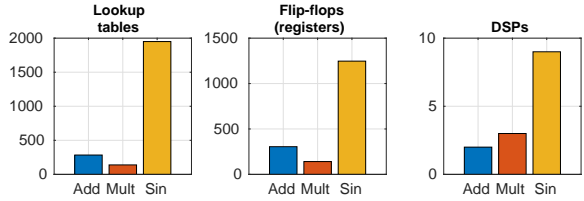} 
	\caption{FPGA resource usage for different computations with single precision floating point arithmetic. Data is obtained from the vendor's high-level synthesis tool, Vivado HLS.}
	\label{fig:fpga_resource}
\end{figure}

\item A vector function $f_c$ is composed of scalar functions that might have different underlying mathematical operations and different evaluation complexities, i.e. $f_c$ might have \textit{irregular} structure. Irregularity potentially limits reusing computational logic and speedup by parallelization.
\end{itemize}

Optimization algorithms, on the other hand, can  benefit from hardware acceleration due to (i)~their iterative nature, which is beneficial for reusing computational logic, and (ii)~the fact that underlying linear algebra algorithms can be efficiently mapped onto hardware~\citep{jerezetalIET:2012}.




Taking the above into account we consider two classes of algorithms for solving~\eqref{cont_nonlin_ocp}: shooting-based and direct transcription algorithms~\citep{betts_nlp}. The common feature of shooting methods is decoupling the ODE and optimization solvers. Accelerating only the latter does not result in significant improvements, due to Amdahl's law, since the workloads of the two operations are comparable. In contrast, direct transcription algorithms transform \eqref{cont_nonlin_ocp} directly to a discrete OCP by approximating the ODE with algebraic equations based on numerical integration equations, i.e. 
\begin{subequations} 
\label{eq:discrete_ocp}
\begin{equation}
  \begin{aligned}
 \min_{\substack{ {x}_0, \dots , {x}_{N} \\ u_0, \dots , u_{N-1} \\ r_0, \dots , r_{N-1}\\  s_0, \dots , s_{N-1}, s_T }} \sum_{k=0}^{N-1} \biggl(T_s ||h(u_k, x_k,s_k)||_2^2 \biggr) + ||h_T(x_N,s_T)||_2^2
 \end{aligned}
\end{equation}
	\begin{align} 
  \textrm{subject to:}  	\\
  			  \quad\quad\quad x_{0} &= \hat{x},   \\
			  \quad c({x}_{k}, u_{k},r_k)&= \begin{bmatrix} x_{k+1} \\ 0 \end{bmatrix} , &&  k = 0,\dots, N -1 \label{eq:colloc_constr}  \\
			  \quad q(x_k,u_k,s_k) &= 0, && k = 0,\dots, N -1   \\
			  \quad q_T(x_N,s_T) &= 0,   \\
			  \quad x_l \leq  x_k  & \leq x_u, && k = 0,\dots, N -1  \\
			  \quad u_l \leq  u_k & \leq u_u, && k = 0,\dots, N -1 \\
		 	  \quad s_l \leq  s_k & \leq s_u, && k = 0,\dots, N -1 \\
		 	  \quad x_{Tl} \leq  x_N & \leq x_{Tu}, \\
		 	  \quad s_{Tl} \leq  s_T & \leq s_{Tu}, \\
	\end{align}
\end{subequations}
\noindent
where $N$ and $T_s$, respectively, denote the horizon length and the sampling time, $ r_k = \left[ r_k^{(1)T}  \ \cdots \  r_k^{(l)T} \right] ^T \in \mathbb{R}^{nl}$ is a vector of integrator intermediate stages and~$l$ is the number of integrator stages per sampling instant. 


With a trapezoidal integrator, \eqref{eq:colloc_constr} would be given by:
\begin{subequations}  \label{trapezoidal_integrator}
	\begin{align}
	 	x_{k}  + T_{s}\left( r_k^{(1)} + r_k^{(2)}\right) /2 &=  x_{k+1} \\
	 	r_k^{(1)} - f_{c}(x_{k}^{}, u_{k})  &= 0 \\
	 	r_k^{(2)} - f_{c}\left( x_{k}^{}+T_{s} \left( r_k^{(1)}+ r_k^{(1)}\right) /2, u_{k}\right)  &= 0 
	 \end{align}
\end{subequations}
\noindent

The  OCP~\eqref{eq:discrete_ocp} can be transformed into an NLP of the  form
\begin{subequations} \label{nlp}
		\begin{align}
		\min_{\theta} \quad   \frac{1}{2} ||f(\theta)||_2^2 \\
		\textrm{subject to} 
		\quad p(\theta) = 0, \\
		J\theta - d\leq 0,
	\end{align}
\end{subequations}
\noindent
where $ \theta = [ x_0^T \ u_0^T \ r_0^T \ s_0^T \dots  x_{N-1}^T \ u_{N-1}^T \ r_{N-1}^T \ s_{N-1}^T \ x_{N}^T \ s_T^{T} ]^T $. Note that $J$ is a matrix of zeros and positive/negative ones since~\eqref{eq:discrete_ocp} contains only bound inequalities.

NLP~\eqref{nlp} incorporates both integration and optimization, which potentially opens the possibility of accelerating both subproblems. Primal-dual interior-point algorithms have proven their efficiency for the numerical solution of optimal control problems~\citep{shahzadetal:2012}. Moreover, with interior-point algorithms, the structure of the KKT matrix associated with the OCP remains fixed (unlike with active set methods), which is desirable for hardware implementations~\citep{jerezetalIET:2012}.

The next section  introduces a structure-exploiting heterogeneous implementation of an interior-point algorithm for solving~\eqref{nlp}. The implementation supports the use of Runge-Kutta family integrators for temporal discretization, considering Butcher tableau as a design parameter. However, results for memory saving, algorithm acceleration and closed loop simulations are obtained with a trapezoidal integrator~\eqref{trapezoidal_integrator}, which belongs to the family of explicit integrators and hence allows efficient handling of stiff ODEs.


%% file: implementation.tex
\section{Algorithm and implementation details}

\subsection{Primal-dual interior-point algorithm}
Interior-point algorithms solve NLPs by incorporating inequality constraints into the objective function using a logarithmic barrier function scaled with a barrier parameter. The modified problem is solved by performing consecutive Newton steps with or without globalization. Algorithm~\ref{alg:ip} and \ref{sec:algorith_details} outline a primal-dual interior-point algorithm for solving~\eqref{nlp}. 
\begin{algorithm}[tb]
	\caption{Primal-dual interior-point method for NLP}	
	\label{alg:ip}
		\begin{algorithmic}[1]
			\State Initial point $[\theta^T_{[0]},\nu^T_{[0]} , \lambda^T_{[0]}]^T: \lambda_{[0]} > 0, G\theta_{[k]}-d < 0$
			\State Reduction parameter $0 < \sigma \leq 1$ 
				\For{$k=0$ to $n_{iter}$}
				\State $A_{[k]} = 
													\begin{bmatrix}
													H_{[k]} + G W_{[k]} G^T & \nabla_{\theta} p^{T}(\theta_{[k]}) \\
													\nabla_{\theta} p(\theta_{[k]}) & 0  \\
												\end{bmatrix}$ 
				\State $b_{[k]} =\begin{bmatrix} r_{dual} \\ r_{eq} \end{bmatrix} $
				\State Solve $A_{[k]}z_{[k]} = b_{[k]}$ for $z_{[k]} = \begin{bmatrix} \Delta \theta_{[k]} \\ \Delta \nu_{[k]}  \end{bmatrix}  $
				
				\State $\Delta \lambda_{[k]} = -\Lambda_{[k]} e - G^{-1} (\theta_{[k]})\mu e  -  G^{-1} (\theta_{[k]}) \Lambda_{[k]} \nabla_{\theta} g(\theta_{[k]}) \Delta \theta_{[k]}$ 
				\State $\alpha_{[k]} = \textrm{max}_{(0,1 \rbrack}\alpha: \lambda_{[k]} + \alpha_{[k]} \Delta \lambda_{[k]} > 0, G(\theta_{[k]} + \Delta \theta_{[k]}) -d  < 0$											
				\State $[\theta^T_{[k+1]},\nu^T_{[k+1]} , \lambda^T_{[k+1]}]^T = \alpha_{[k]}[\Delta \theta^T_{[k]},\Delta  \nu^T_{[k]} ,\Delta  \lambda^T_{[k]}]^T + [ \theta^T_{[k]},  \nu^T_{[k]} ,  \lambda^T_{[k]}]^T$
			\EndFor
		\end{algorithmic}
	\end {algorithm}
The algorithm computes a barrier parameter based on the current complementary value and reduction parameter, which is one of the simplest and most common approaches~\citep{barrier_ip}. Furthermore, there is no globalization strategy incorporated in the algorithm. This choice is justified by the fact that line search and trust region globalization algorithms involve repetitive evaluation of $f(\theta)$ and $p(\theta)$, which, as was previously mentioned, often have irregular structure and hence discourage efficient acceleration. An extensive comparative study of barrier update strategies and globalization techniques for nonlinear interior-point methods can be found in~\citep{barrier_ip}.

Another algorithmic choice that requires justification is the Hessian approximation method. Most of existing interior-point algorithms for large scale optimization use either Broyden-Fletcher-Goldfarb-Shanno (BFGS)~\cite{bfgs} or Gauss-Newton approximators. The BFGS algorithm, if applied blockwise, can achieve a block diagonal structure for the Hessian approximation matrix when used for optimal control applications~\cite{block_bfgs}. The Gauss-Newton algorithm, in addition, preserves sparsity within each block, which is particularly beneficial taking into account scratchpad memory limitations of reconfigurable computers. However, memory savings come at the price of narrowing the application scope to least-squares types of problems. In this work we accept this limitation and use the Gauss-Newton approach.

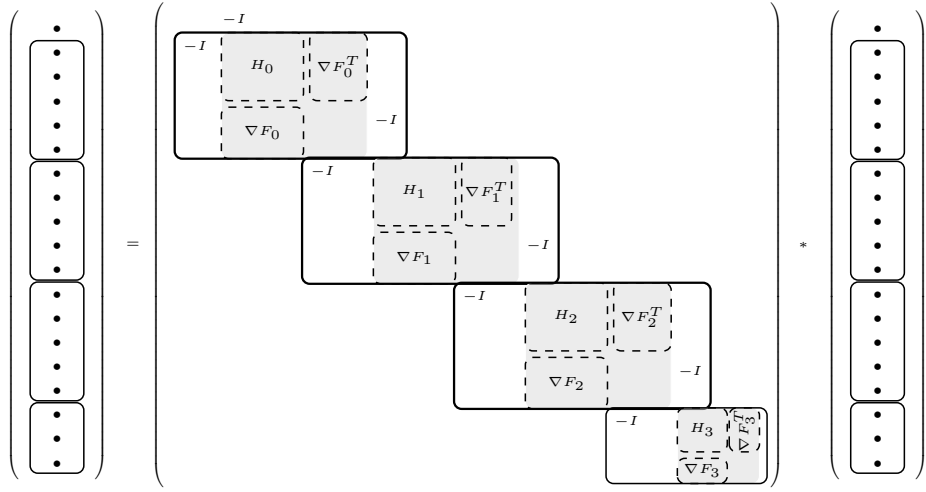
\begin{figure}[t!]
	\centering
	\input{kkt_structure.tex}
	\caption{Structure of the KKT matrix assosiated with the problem~\eqref{nlp}. $H_i$ and $\nabla F_i$ denote Hessians and Jacobians of the corresponding sampling instant in the prediction horizon. }
	\label{fig:matrix_structure}
\end{figure}

The main  workload of the algorithm is concentrated in solving the system of linear equations. The matrix associated with the problem is symmetric and can be reordered to achieve the structure presented in Figure~\ref{fig:matrix_structure}.
A symmetric system of linear equations can be solved using direct methods, e.g.\ LDL decomposition, or iterative methods, e.g.\ Minimum Residual (MINRES) algorithm~\citep{minres}. Decomposition algorithms often involve many division computations, which are more complicated from a hardware implementation point of view compared to addition and multiplication. Moreover, parallelizing computations is not straightforward with direct algorithms. In contrast, iterative methods mainly rely on matrix-vector multiplications, while having very few division and square root computations. In this work we use the MINRES algorithm, which can be efficiently mapped onto hardware~\citep{minres_fpga} and is well studied in relation to optimal control applications~\citep{shahzadetal:2012}. The algorithm performs minimization of the residual $||b - Az||_{2}$ over a Krylov subspace, which is constructed iteratively by the Lanczos kernel~\citep{minres}. The Lanczos kernel is based on a three-term recurrence, hence there is no need to store the entire Krylov subspace (unlike with the generalized minimal residual method~\citep{gmres}). From an implementation point of view, the MINRES algorithm is mostly based on addition and multiplication operations, requiring only two scalar divisions and two scalar square root computations per iteration.

\subsection{Proposed implementation}

\begin{figure}[tb]
	\centering
	\includegraphics[width=0.6\textwidth]{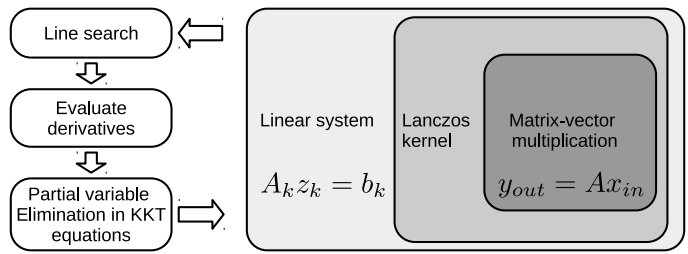} 
	\caption{Algorithm~\ref{alg:ip} flow diagram with MINRES solver.}
	\label{fig:algorithm_flow}
\end{figure}

Algorithm~\ref{alg:ip} with the MINRES linear solver can be visualized with the block diagram in Figure~\ref{fig:algorithm_flow}.
We consider four different ways of splitting the workload between software and hardware:
\begin{itemize}
\item The entire algorithm is implemented in software, i.e. ARM Cortex-A9 processor. We denote this implementation by SW.
\item Only the matrix-vector multiplication is accelerated in hardware and the rest is implemented in software (HG$_1$).
\item The whole Lanczos kernel is accelerated in hardware and the rest is implemented in software (HG$_2$).
\item The whole linear system solver is accelerated in hardware and the rest is implemented in software (HG$_3$).
\end{itemize}  

\begin{figure}[tb]
	\centering
	\includegraphics[width=0.6\textwidth]{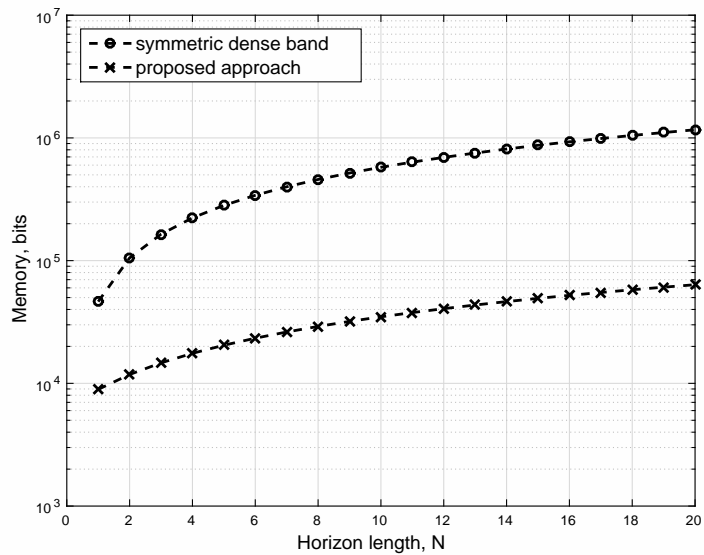} 
	\caption{Memory requirements for the storing matrix on Figure~\ref{fig:matrix_structure}. ODE model and OCP parameters are presented in Sections~\ref{sec:experimental_setup} and~\ref{sec:experimental_results}, trapezoidal method is used for ODE integration. Details on dense band implementation can be found in~\citep{boland_memory_bandwidth}. Actual memory savings may vary depending on the memory block size of a particular computing platform.}
	\label{fig:matrix_memory}
\end{figure}

Note that for all considered options, the Jacobians evaluations are implemented in software to avoid synthesising resource-consuming nonlinear operators. Operations that are implemented in hardware  can be classified into:
\begin{itemize}
\item \textit{Scalar operations}, which do not require acceleration. 
\item \textit{Vector-vector operations}, which can be efficiently pipelined in hardware.
\item \textit{Matrix-vector multiplication}, which is the most resource-consuming part. Efficient implementation requires exploiting sparsity.
\end{itemize}
One possible way to exploit the structure of $A_k$  is based on considering the matrix as banded~\citep{jerezetalIET:2012}. In this case the number of parallel computations is defined by the bandwidth and cannot be changed with respect to resource availability. In this work we treat the matrix as block sparse and, in addition, exploit sparsity within each block, which can result in an 18x saving in memory (Figure~\ref{fig:matrix_memory}).

Matrix-vector multiplication can potentially be parallelized by simultaneous processing of the  solid line blocks in Figure~\ref{fig:matrix_structure}, which correspond to different sampling instants. However, consecutive blocks are coupled, i.e.\ each block accesses areas in the input vector associated with its neighbours, which results in restrictions on memory partitioning. It can be noted that blocks are coupled only via negative identity matrices. For a matrix-vector multiplication, handling negative identity matrices is reduced to data transfer operations (with changing the sign bit) and does not require any arithmetic operators. After negative identity matrices are processed, the remaining parts (grey area) can be parallelized, since there is no overlap in accessing input vector partitions.
 
\begin{figure}[tb]
\centering
\begin{subfigure}[b]{0.45\textwidth}
\includegraphics[width=\linewidth,keepaspectratio=true]{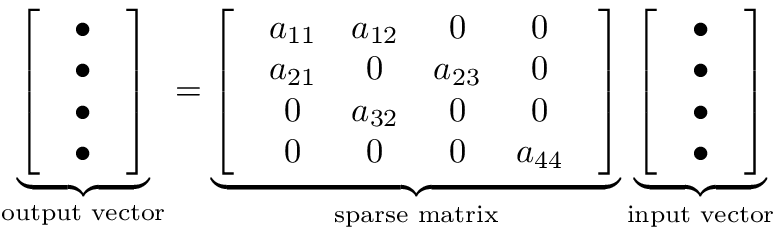}
\caption{}
\end{subfigure}
\begin{subfigure}[b]{0.7\textwidth}
\includegraphics[width=\linewidth,keepaspectratio=true]{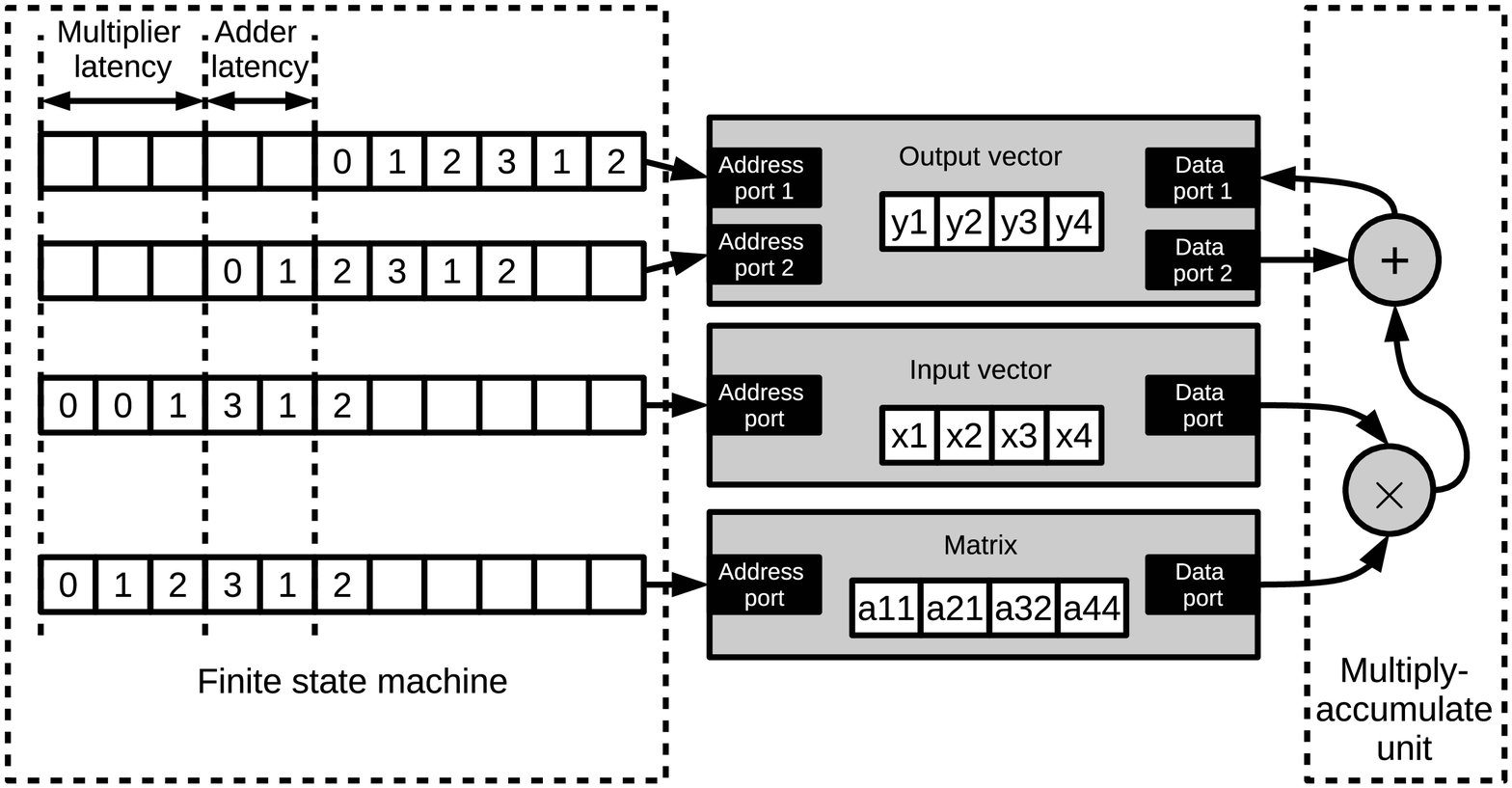}
\caption{}
\end{subfigure}
\caption{An example of a sparse matrix (a) and the corresponding circuit (b) for matrix vector multiplication.}
\label{fig:block_toy_example}
\end{figure}

The efficiency of multiplication  also depends on how sparsity within each block is handled. For example, consider the sparse matrix in Figure~\ref{fig:block_toy_example}. 
Each non-zero element in the matrix corresponds to a multiply-accumulate (MAC) operation, i.e.\ each non-zero element is multiplied by a corresponding element of the input vector and added to the corresponding element of the output vector.  Although the order of processing non-zero elements with floating point arithmetic might affect the final result, in practice this effect is assumed to be negligible. More importantly, this order has an impact on   output vector read-write dependencies and hence pipelining possibilities. For the considered example, processing of $a_{12}$ cannot be started before $a_{11}$ is fully processed, since both elements require writing data to the same element of the output vector. Hence, MAC operations have to be scheduled in such a way that the distance (in terms of computer clock cycles) between processing non-zero elements from the same row is maximized. This scheduling problem can be formulated as an optimization problem
\begin{subequations} 
\label{eq:scheduling}
\begin{equation} 
  \begin{aligned}
	\max_{d, S_{sched}} \quad  d
 \end{aligned}
\end{equation}
subject to:
	\begin{align} 
	& |t(a_{ij}) - t(a_{kl})|  > d  && \forall a_{ij} \in M, \forall a_{kl} \in M: i=k, j \neq l \\
	 & 1 \leq t(a_{ij}) \leq N_{nz} && \forall a_{ij} \in M \\
	& t(a_{ij})  \neq t(a_{kl}) && \iff i \neq k, j \neq l 
	\end{align}
\end{subequations}
\noindent
where $d \in \mathbb{Z}$ is a slack variable, $M=\{a_{11},$ $a_{12},$ $a_{21},$ $a_{23},$ $a_{32},$ $a_{44} \}$ is the set of non-zero elements, $N_{nz}$ is the number of non-zero elements, $t(a_{ij})$ is the number of time instant, at which processing of $a_{ij}$ starts, and $S_{sched} = \lbrack t(a_{11}),$ $t(a_{12}),$ $t(a_{21}),$ $t(a_{23}),$ $t(a_{32}),$ $t(a_{44}) \rbrack^T \in \mathbb{Z}^{N_{nz}}$. Since all blocks in Figure~\ref{fig:matrix_structure} have the same sparsity pattern, the scheduling problem is solved only once during design. The resulting MAC circuit for the considered example is shown in Figure~\ref{fig:block_toy_example}. Note that due to symmetry of $A_k$ only the lower triangular part is stored.

\begin{figure}[tb]
	\centering
	\includegraphics[width=0.6\textwidth]{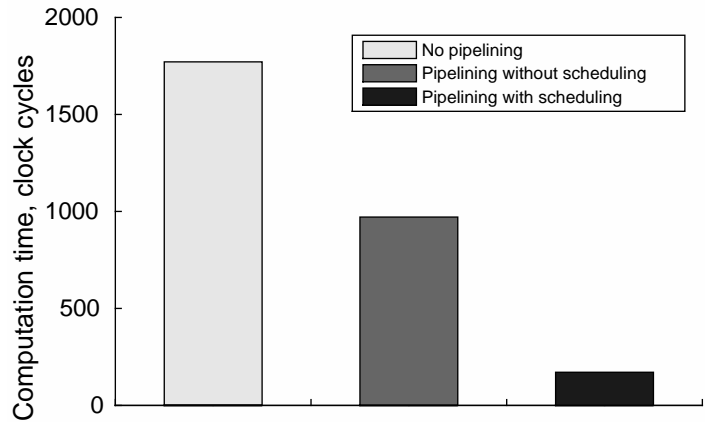} 
	\caption{The impact of scheduling on computation time of a sparse matrix-vector multiplication (for each gray block in Figure~\ref{fig:matrix_structure}). It is assumed that the latency for the addition and multiplication units  are 6 and 5 clock cycles, respectively. The number of nonzero elements is 161, which is the case for the example described in Sections~\ref{sec:experimental_setup} and~\ref{sec:experimental_results}.}
	\label{fig:scheduling_effect}
\end{figure}

The impact of scheduling on the time taken for sparse matrix-vector multiplication  is presented in Figure~\ref{fig:scheduling_effect}. 
For this case study, problem~\eqref{eq:scheduling} was solved using the YALMIP built-in branch and bound solver~\citep{yalmip}. If pipelining is implemented without scheduling, the \textit{initiation interval} is equal to the adder latency, which limits potential improvement. Initiation interval is defined as the number of clock cycles that elapses between starting processing two successive non-zeros. With a systematic scheduling approach, read-write dependencies are handled optimally, which allows one to start processing a new non-zero element after the previous as quickly as possible, i.e.\ achieving the minimal initiation interval. For the ODE model presented in Section~\ref{sec:experimental_setup} it was possible to achieve initiation interval of one clock cycle.

\begin{figure}[tb]
	\centering
	\includegraphics[width=0.6\textwidth]{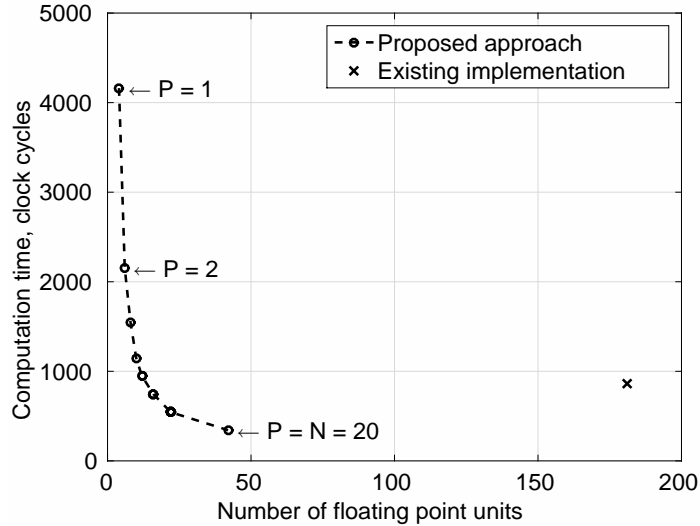} 
	\caption{Trading off computational time against processor resource usage for matrix-vector multiplication (Figure~\ref{fig:matrix_structure}). Each circle represents a design with a certain degree of parallelism. The floating point unit is either floating point addition or multiplication. For the considered test case, $N = 20$ and the ODE model is given in Section~\ref{sec:experimental_setup}.  It is assumed that the latency of the addition and multiplication units  are 6 and 5 clock cycles, respectively. Details on the existing implementation can be found in~\citet{boland_memory_bandwidth}.}
	\label{fig:trade_off}
\end{figure}

Although matrix-vector multiplication can be parallelized by allocating a MAC unit to each grey block (Figure~\ref{fig:matrix_structure}), we propose trading off computation time against resource usage by varying the number of MAC units, as shown in Figure~\ref{fig:trade_off}. The number of MAC units is denoted by $P$.

\subsection{Limitations of the proposed implementation}

We highlight certain limitations of the proposed implementation:

\begin{itemize}
	\item For matrix-vector multiplication, input and output vectors have to be partitioned in accordance with the matrix Figure~\ref{fig:matrix_structure}, which, depending on the parallelism level $P$, might increase FPGA block memory usage. This issue can be partially addressed by placing corresponding partitions of the matrix and input vector in the same memory block, which will not affect performance, provided the computing platforms supports dual-port memory blocks. Another possible solution is implementing vector partitions using lookup tables (provided partitions are sufficiently small) instead of using block memory.
	\item In contrast to~\citep{boland_memory_bandwidth}, our implementation requires control logic, which might limit FPGA clock frequency. However, matrix-vector multiplication, being a part of a larger algorithm, is often not the main bottleneck for increasing clock frequency~\citep{minres_fpga}.
	\item Applying a preconditioner directly to original linear system may destroy in-block sparsity and hence eliminate memory savings. To overcome this limitation, preconditioners can be applied indirectly by performing two matrix-vector multiplications per MINRES iteration so that sparsity patterns are preserved, both for preconditioner and the original matrix~\cite{greenbaum_iterative_methods}. In this work we use a sparsity-preserving prescaler described in~\citep{lanczos_scaling}. Although preserving sparsity, this prescaler changes negative identity matrices into general diagonal matrices, which slightly complicates the algorithm.
\end{itemize}

%% file: kkt_structure.tex
\begin{tikzpicture}
    \tiny
        \matrix (m) [matrix of math nodes,  left delimiter=(,right delimiter=),inner sep=1mm] 
        {
			\ws&\nI&\ws&\ws&\ws&\ws&\ws&\ws&\ws&\ws&\ws&\ws&\ws&\ws&\ws&\ws&\ws&\ws&\ws \\   
			\nI&\ws&\ws&\ws&\ws&\ws&\ws&\ws&\ws&\ws&\ws&\ws&\ws&\ws&\ws&\ws&\ws&\ws&\ws \\  
			\ws&\ws&\ws&\ws&\ws&\ws&\ws&\ws&\ws&\ws&\ws&\ws&\ws&\ws&\ws&\ws&\ws&\ws&\ws \\ 
			\ws&\ws&\ws&\ws&\ws&\ws&\ws&\ws&\ws&\ws&\ws&\ws&\ws&\ws&\ws&\ws&\ws&\ws&\ws \\   
			\ws&\ws&\ws&\ws&\ws&\ws&\nI&\ws&\ws&\ws&\ws&\ws&\ws&\ws&\ws&\ws&\ws&\ws&\ws \\  
			\ws&\ws&\ws&\ws&\ws&\ws&\ws&\ws&\ws&\ws&\ws&\ws&\ws&\ws&\ws&\ws&\ws&\ws&\ws \\   
			\ws&\ws&\ws&\ws&\nI&\ws&\ws&\ws&\ws&\ws&\ws&\ws&\ws&\ws&\ws&\ws&\ws&\ws&\ws \\   
			\ws&\ws&\ws&\ws&\ws&\ws&\ws&\ws&\ws&\ws&\ws&\ws&\ws&\ws&\ws&\ws&\ws&\ws&\ws \\ 
			\ws&\ws&\ws&\ws&\ws&\ws&\ws&\ws&\ws&\ws&\ws&\ws&\ws&\ws&\ws&\ws&\ws&\ws&\ws \\        
			\ws&\ws&\ws&\ws&\ws&\ws&\ws&\ws&\ws&\ws&\ws&\nI&\ws&\ws&\ws&\ws&\ws&\ws&\ws \\ 
			\ws&\ws&\ws&\ws&\ws&\ws&\ws&\ws&\ws&\ws&\ws&\ws&\ws&\ws&\ws&\ws&\ws&\ws&\ws \\
			\ws&\ws&\ws&\ws&\ws&\ws&\ws&\ws&\ws&\nI&\ws&\ws&\ws&\ws&\ws&\ws&\ws&\ws&\ws \\  
			\ws&\ws&\ws&\ws&\ws&\ws&\ws&\ws&\ws&\ws&\ws&\ws&\ws&\ws&\ws&\ws&\ws&\ws&\ws \\
			\ws&\ws&\ws&\ws&\ws&\ws&\ws&\ws&\ws&\ws&\ws&\ws&\ws&\ws&\ws&\ws&\ws&\ws&\ws \\
			\ws&\ws&\ws&\ws&\ws&\ws&\ws&\ws&\ws&\ws&\ws&\ws&\ws&\ws&\ws&\ws&\nI&\ws&\ws \\
			\ws&\ws&\ws&\ws&\ws&\ws&\ws&\ws&\ws&\ws&\ws&\ws&\ws&\ws&\ws&\ws&\ws&\ws&\ws \\
			\ws&\ws&\ws&\ws&\ws&\ws&\ws&\ws&\ws&\ws&\ws&\ws&\ws&\ws&\nI&\ws&\ws&\ws&\ws \\ 
			\ws&\ws&\ws&\ws&\ws&\ws&\ws&\ws&\ws&\ws&\ws&\ws&\ws&\ws&\ws&\ws&\ws&\ws&\ws \\    
			\ws&\ws&\ws&\ws&\ws&\ws&\ws&\ws&\ws&\ws&\ws&\ws&\ws&\ws&\ws&\ws&\ws&\ws&\ws \\      
        };
        \draw [color=white,rounded corners=0.1cm, fill=black!15, fill opacity=0.5] (m-6-2.south west) rectangle (m-2-6.north east) node[pos=.5] {$$};
        \draw[color=black,rounded corners=0.1cm, dashed,line width=0.2mm] ([shift={(0mm,1mm)}]m-4-2.south west) rectangle (m-2-4.north east) node[pos=.5] {$H_0$};
        \draw[color=black,rounded corners=0.1cm, dashed,line width=0.2mm] (m-6-2.south west) rectangle (m-5-4.north east) node[pos=.5] {$\nabla F_0$};
        \draw[color=black,rounded corners=0.1cm, dashed,line width=0.2mm] ([shift={(0mm,1mm)}]m-4-5.south west) rectangle (m-2-6.north east) node[pos=.5] {$\nabla F^T_0$};
        \draw[color=black,rounded corners=0.1cm,line width=0.3mm] ([shift={(-1mm,0mm)}]m-6-1.south west) rectangle ([shift={(1mm,0mm)}]m-2-7.north east) node[pos=.5] {$$};

        \draw [color=white,rounded corners=0.1cm, fill=black!15, fill opacity=0.5] (m-11-7.south west) rectangle (m-7-11.north east) node[pos=.5] {$$};
        \draw[color=black,rounded corners=0.1cm, dashed,line width=0.2mm] ([shift={(0mm,1mm)}]m-9-7.south west) rectangle (m-7-9.north east) node[pos=.5] {$H_1$};
        \draw[color=black,rounded corners=0.1cm, dashed,line width=0.2mm] (m-11-7.south west) rectangle (m-10-9.north east) node[pos=.5] {$\nabla F_1$};
        \draw[color=black,rounded corners=0.1cm, dashed,line width=0.2mm] ([shift={(0mm,1mm)}]m-9-10.south west) rectangle ([shift={(-1mm,0mm)}]m-7-11.north east) node[pos=.5] {$\nabla F^T_1$};
     \draw[color=black,rounded corners=0.1cm,line width=0.3mm] ([shift={(-1mm,0mm)}]m-11-5.south west) rectangle ([shift={(1mm,0mm)}]m-7-12.north east) node[pos=.5] {$$};

        \draw [color=white,rounded corners=0.1cm, fill=black!15, fill opacity=0.5] (m-16-12.south west) rectangle (m-12-16.north east) node[pos=.5] {$$};
        \draw[color=black,rounded corners=0.1cm, dashed,line width=0.2mm] ([shift={(0mm,1mm)}]m-14-12.south west) rectangle (m-12-14.north east) node[pos=.5] {$H_2$};
        \draw[color=black,rounded corners=0.1cm, dashed,line width=0.2mm] (m-16-12.south west) rectangle (m-15-14.north east) node[pos=.5] {$\nabla F_2$};
        \draw[color=black,rounded corners=0.1cm, dashed,line width=0.2mm] ([shift={(0mm,1mm)}]m-14-15.south west) rectangle (m-12-16.north east) node[pos=.5] {$\nabla F^T_2$};
    \draw[color=black,rounded corners=0.1cm,line width=0.3mm] ([shift={(-1mm,0mm)}]m-16-10.south west) rectangle ([shift={(1mm,0mm)}]m-12-17.north east) node[pos=.5] {$$};       
    
        \draw [color=white,rounded corners=0.1cm, fill=black!15, fill opacity=0.5] (m-19-17.south west) rectangle (m-17-19.north east) node[pos=.5] {$$};
        \draw[color=black,rounded corners=0.1cm, dashed,line width=0.2mm] ([shift={(0mm,1mm)}]m-18-17.south west) rectangle ([shift={(-1mm,0mm)}]m-17-18.north east) node[pos=.5] {$H_3$};
        \draw[color=black,rounded corners=0.1cm, dashed,line width=0.2mm] (m-19-17.south west) rectangle ([shift={(-1mm,0mm)}]m-19-18.north east) node[pos=.5] {$\nabla F_3$};
        \draw[color=black,rounded corners=0.1cm, dashed,line width=0.2mm] ([shift={(-0.6mm,1mm)}]m-18-19.south west) rectangle (m-17-19.north east) node[pos=.5] {$\rotatebox{90}{\text{$\nabla F^T_3$}}$};
   \draw[color=black,rounded corners=0.1cm,line width=0.2mm] ([shift={(-1mm,0mm)}]m-19-15.south west) rectangle ([shift={(1mm,0mm)}]m-17-19.north east) node[pos=.5] {$$};
   	\matrix (v_in) [matrix of math nodes,  left delimiter=(,right delimiter=),inner sep=1mm, right=1cm of m]
	{
	 \ws \bullet \ws \\ \ws \bullet \ws \\  \ws \bullet \ws \\  \ws \bullet \ws \\  \ws \bullet \ws \\   
	 \ws \bullet \ws \\ \ws \bullet \ws \\  \ws \bullet \ws \\  \ws \bullet \ws \\  \ws \bullet \ws \\  
	 \ws \bullet \ws \\ \ws \bullet \ws \\  \ws \bullet \ws \\  \ws \bullet \ws \\  \ws \bullet \ws \\  
	 \ws \bullet \ws \\ \ws \bullet \ws \\  \ws \bullet \ws \\  \ws \bullet \ws \\  
	};   
	\draw[color=black,rounded corners=0.1cm, solid,line width=0.2mm] ([shift={(0mm,0.2mm)}]v_in-6-1.south west) rectangle ([shift={(0mm,-0.2mm)}]v_in-2-1.north east) node[pos=.5] {};
	\draw[color=black,rounded corners=0.1cm, solid,line width=0.2mm] ([shift={(0mm,0.2mm)}]v_in-11-1.south west) rectangle ([shift={(0mm,-0.2mm)}]v_in-7-1.north east) node[pos=.5] {};
	\draw[color=black,rounded corners=0.1cm, solid,line width=0.2mm] ([shift={(0mm,0.2mm)}]v_in-16-1.south west) rectangle ([shift={(0mm,-0.2mm)}]v_in-12-1.north east) node[pos=.5] {};
	\draw[color=black,rounded corners=0.1cm, solid,line width=0.2mm] ([shift={(0mm,0.2mm)}]v_in-19-1.south west) rectangle ([shift={(0mm,-0.2mm)}]v_in-17-1.north east) node[pos=.5] {};
   	\matrix (v_out) [matrix of math nodes,  left delimiter=(,right delimiter=),inner sep=1mm, left=1cm of m]
	{
	 \ws \bullet \ws \\ \ws \bullet \ws \\  \ws \bullet \ws \\  \ws \bullet \ws \\  \ws \bullet \ws \\   
	 \ws \bullet \ws \\ \ws \bullet \ws \\  \ws \bullet \ws \\  \ws \bullet \ws \\  \ws \bullet \ws \\  
	 \ws \bullet \ws \\ \ws \bullet \ws \\  \ws \bullet \ws \\  \ws \bullet \ws \\  \ws \bullet \ws \\  
	 \ws \bullet \ws \\ \ws \bullet \ws \\  \ws \bullet \ws \\  \ws \bullet \ws \\  
	};   
	\draw[color=black,rounded corners=0.1cm, solid,line width=0.2mm] ([shift={(0mm,0.2mm)}]v_out-6-1.south west) rectangle ([shift={(0mm,-0.2mm)}]v_out-2-1.north east) node[pos=.5] {};
	\draw[color=black,rounded corners=0.1cm, solid,line width=0.2mm] ([shift={(0mm,0.2mm)}]v_out-11-1.south west) rectangle ([shift={(0mm,-0.2mm)}]v_out-7-1.north east) node[pos=.5] {};
	\draw[color=black,rounded corners=0.1cm, solid,line width=0.2mm] ([shift={(0mm,0.2mm)}]v_out-16-1.south west) rectangle ([shift={(0mm,-0.2mm)}]v_out-12-1.north east) node[pos=.5] {};
	\draw[color=black,rounded corners=0.1cm, solid,line width=0.2mm] ([shift={(0mm,0.2mm)}]v_out-19-1.south west) rectangle ([shift={(0mm,-0.2mm)}]v_out-17-1.north east) node[pos=.5] {};
    \matrix (mult) [matrix of math nodes,inner sep=1mm, left=0.25cm of v_in]
	{
	 *\\  
	};
    \matrix (mult) [matrix of math nodes,inner sep=1mm, right=0.25cm of v_out]
	{
	 =\\  
	};   
    \end{tikzpicture}

%% file: experimental_setup.tex
\section{Experimental setup}

\label{sec:experimental_setup}

Experimental setup represents a closed-loop system that consists of a gantry crane model and a heterogeneous computer running the predictive controller. The closed-loop simulation is managed by Protoip, a software tool for rapid prototyping of online optimization algorithms on reconfigurable computing platforms. This section will provide detailed information on the experimental setup, including a new release of Protoip, heterogeneous computer and the gantry crane model.

\subsection{Protoip - a new release}

\begin{figure}
    \centering
    \begin{minipage}{0.5\textwidth}
        \centering
        \includegraphics[width=0.95\textwidth]{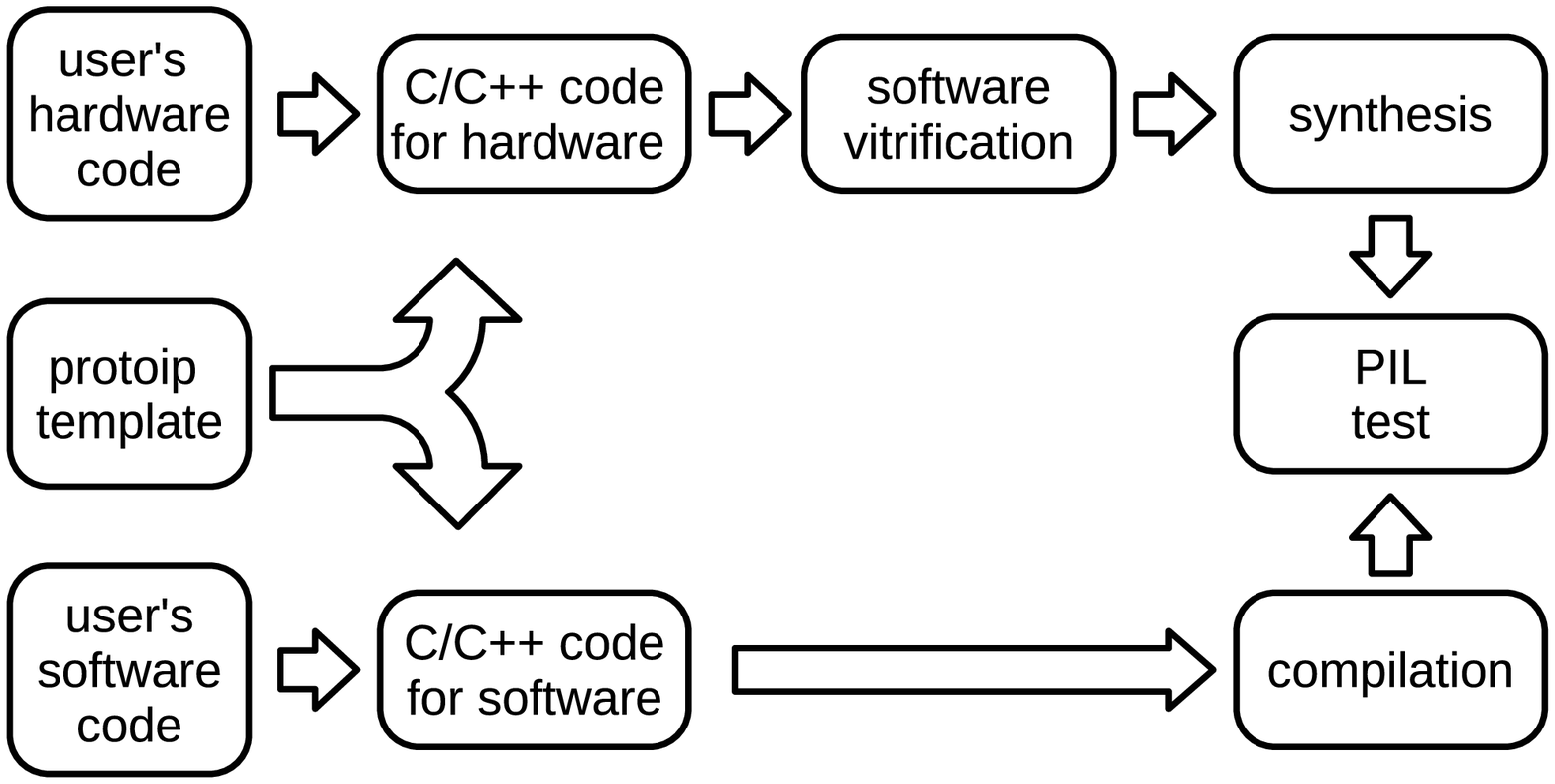}
        \caption{SoC design flow with Protoip.}
        \label{fig:protoip_flow}
    \end{minipage}\hfill
	\begin{minipage}{0.5\textwidth}
        \centering
        \includegraphics[width=0.95\textwidth]{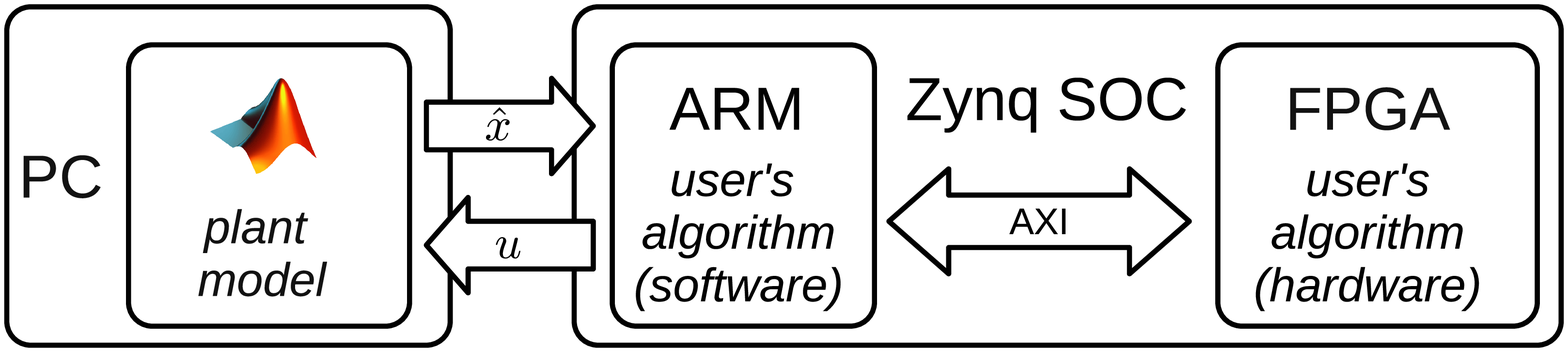}
        \caption{Processor-in-the-loop test with Protoip.}
        \label{fig:protoip_pil}
    \end{minipage}
\end{figure}

The initial release of Protoip~\citep{protoip} represents a software tool for rapid deploying and verification of online optimization algorithms on FPGAs using hardware vendor's tools on the underlying level (Xilinx Vivado, Vivado HLS, SDK). Protoip is written in Tcl, an interpreted programming language that was designed with an emphasis on computer aided design. Matlab interface is provided to allow working in conjunction with state-of-the-art optimization and numerical integration toolboxes. The previous version of Protoip was dealing only with pure FPGA implementations, i.e. CPU was not involved in computations. The new release adds possibility of splitting the workload between CPU and FPGA within a single chip to exploit strengths of each subsystem and trade off FPGA logic usage against algorithm performance. The design flow for heterogeneous implementations consists of two parallel branches which correspond to software and hardware parts of a given algorithm accordingly (Figure~\ref{fig:protoip_flow}). 
Initially, the user provides two pieces of C/C++ code: software code for CPU and hardware code for FPGA. This step can be performed either manually or automated by using synthesizable code generation tools~\citep{protosplit}. On the next stage, Protoip performs software verification of the hardware code using automatically generated testbench files, which allows detecting errors on the early stages of the design flow avoiding time consuming circuit synthesis process. Following this, FPGA circuit is synthesised and software code is compiled. On the final stage of the design flow implementation is deployed on a heterogeneous platform and Processor-in-the-loop (PIL) test is performed.

Processor-in-the-loop test implies simulating the controlled plant in Matlab environment on PC, while performing computations on an embedded platform (Figure~\ref{fig:protoip_pil}). Desktop PC to embedded platform communication speed and reliability might be critical for processor- in-the-loop simulations. The previous release of Protoip was relying on UPD/Ethernet interface, which is a common and relatively fast protocol, however, having no transmission guarantees. Lack of reliability might be crucial for design space space exploration, where processor-in- the-loop simulations may run autonomously for several weeks. A possible solution to this problem is employing PCI interface, which has proven to be reliable solution for FPGA to PC communication~\cite{fpga_pci}. However, PCI interface is rarely available on low-cost prototyping boards. For that reason it was decided to keep using UDP/Ethernet so that no additional hardware is introduced. To improve reliability another communication layer with error checking and retransmission mechanism was implemented on the top of UDP. As a result communication reliability was improved without changing hardware requirements.

Another feature of the release is a new way of managing Vivado, Vivado HLS and Xilinx SDK projects. In particular:
\begin{itemize}
  	  		\item Users get flexibility of managing projects (e.g.\ adding new libraries, project parameters). With the previous version of Protoip only predefined set of project parameters was supported. 
  		\item A possibility of handling multiple file C/C++ projects is added. This make Protoip more friendly in relation to code generation tools. Consider~\citep{protosplit}, a C code generation tool for model predictive control based on operator splitting methods, which uses Protoip on the underlying level and is capable generating code for FPGAs.
  	\end{itemize}

Summary of the new Protoip release features, i.e. contributions of this work:
\begin{itemize}
	\item Possibility of prototyping heterogeneous implementations.
	\item A new communication layer on the top of UDP/Ethernet for reliable processor-in-the-loop simulations.
	\item A new way of managing application projects to allow working in conjunction with code generation tools.
\end{itemize}

\subsection{Target heterogeneous computer}

Protoip supports the entire family of Xilinx Zynq-7000 devices. The proposed implementation was tested on a Xilinx Zynq-7000 XC7Z020 SoC with dual-core ARM Cortex-A9 processor (only one core was used in this work) and Artix-7 FPGA logic that contains 53200 six-input Lookup Tables (LUTs) and 106400 Flip-Flops (FFs). In addition to general purpose logic the Artix-7 provides special purpose units: 220 DSP blocks and 140 block RAMs with total capacity 4.9\,Mb, which is a relatively small amount compared to an average modern CPU RAM. Communication between software and hardware was performed via Advanced eXtensible Interface (AXI) with burst mode support that allows reading/writing of words every clock cycle. We used the Zedboard development kit~\citep{zedboard}, which supplements the Zynq platform with auxilary circuits to provide an easy access to the communication interfaces, including JTAG for programming FPGA and Ethernet for data exchange.

\subsection{Gantry crane}
\begin{figure}[tb]
	\centering
	\includegraphics[width=0.4\columnwidth]{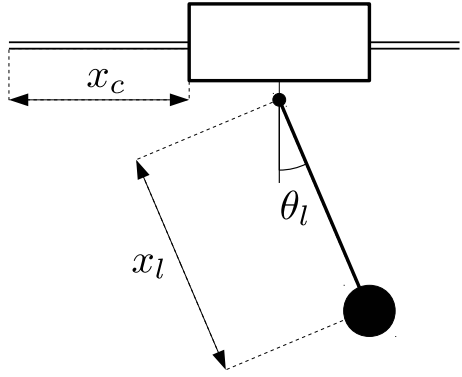} 
	\caption{A schematic drawing of the gantry crane setup.}
	\label{fig:crane_setup}
\end{figure}

For the purpose of closed-loop simulations, Simulink Multibody library-based~\citep{matlab_bib} gantry crane model was used, which allowed incorporation of various types of friction including  viscous, breakaway and Coulomb frictions. The setup consists of a payload ($m$ = 0.47 kg) hanging on a rope attached to a moving cart (Figure~\ref{fig:crane_setup}).
The system is actuated by two Pulse Width Modulation (PWM) controlled motors, which move the cart along the rail and lift/lower the load. Motors are equipped with internal velocity controllers, hence the inputs to the system are velocity setpoints. Cart position $x_c$, rope length $x_l$ and deflection angle $\theta_l$ are measured by encoders with a resolution equal to 4096 pulses per rotation. Corresponding velocities are estimated using a second order low-pass filter cascaded with a differentiator with Laplace transform\noindent
\begin{align}
Y_{est}(s) = \frac{s\omega^{2}}{s^2 + 2\omega \zeta + \omega^{2}}. \label{filter_est}
\end{align}

In this experiment $\omega = 20\pi \textrm{\thinspace rad/s}$, $\zeta = 0.7$. For the purpose of digital implementation~\eqref{filter_est} is discretized using a Tustin transformation with sampling time $T_{est} = 0.01\textrm{s}$. Note that estimator and controller can be sampled at different rates.

The motion of the crane with a variable cable length can be described by~\citep{crane_model}
\begin{equation}
\ddot \theta_l x_l + 2 \dot x_l \dot \theta_l +  \ddot x_c \cos ( \theta_l) + g \sin (\theta_l) = 0,
\end{equation}
\noindent
where $g$ is the acceleration of gravity.

It is assumed that the dynamics between the motor velocity setpoints and corresponding velocities are described by the first-order lag relation with Laplace transform 
\begin{equation}
Y_{motor}(s) = \frac{1}{\tau s +1},
\end{equation}
\noindent
where $\tau$ is the time constant.
The following state-space model can be obtained:
\begin{subequations}
\begin{align}
\dot x_c &= v_c  \\
\dot v_c &= \frac{-v_c+u_c}{\tau_c} \\
\dot x_l &= v_l  \\
\dot v_l &= \frac{-v_l+u_l}{\tau_l} \\
\dot \theta_l &= \theta_l \\
\ddot \theta_l &= \frac{1}{x_l} \left( \frac{-v_c+u_c}{\tau_c}\cos (\theta_l) +g \sin(\theta_l) + 2 v_l \dot \theta_l \right) 
\end{align}
\end{subequations}
\noindent
where $v_c$ and $v_l$ denote corresponding velocities. Time constants $\tau_c = 0.13s$ and $\tau_l = 0.07s$ were identified using sum-of-sinusoids input signal postprocessing data with the nonlinear least squares algorithm from~\citep{matlab_bib}.

In this work controller performance was initially verified by using the same model for predictive control and plant simulation adopting the CasADi~\citep{casadi} front-end of the CVODES numerical integrator~\citep{sundials}. Following initial verification a heterogeneous MPC controller was tested in the loop with Simscape model within the Simulink environment. Incorporating the Zynq platform into the Simulink environment was performed using the Protoip Application Programming Interface (API), which is implemented in C, and the Simulink Coder package, which allows calling external C/C++ functions.

%% file: experimental_results.tex
\section{Experimental results}

\label{sec:experimental_results}

Four implementations (software only and three heterogeneous) of Algorithm~\ref{alg:ip} were deployed and tested with Protoip. Single precision floating point data arithmetic was used, clocking the CPU and FPGA  at 667\,MHz and 167\,MHz, respectively.
The following control objectives were selected:
\begin{equation} \label{eq:objective}
	h(u,x,s) = \begin{pmatrix} x_c + x_l \sin (\theta_l) \\
											x_l \cos (\theta_l) -0.5 \\
											\dot \theta_l \\
											10^{-4} u_c \\
											10^{-4} u_l \end{pmatrix}, \quad
							h_T(x_N,s_T) = \begin{pmatrix} x_c + x_l \sin 													(\theta_l) \\
											x_l \cos (\theta_l) - 0.5 \\
											\dot \theta_l \end{pmatrix}.
\end{equation}
Note that this formulation has proven to be a tuning-free solution for gantry crane control, in contrast to the standard quadratic objective approach, which often relies on a time consuming tuning process~\cite{acado_crane}. Actuation limits dictated by the motor characteristics were incorporated as input constraints:
\begin{subequations} \label{eq:constraints}
\begin{equation}
	-0.15 \leq u_c \leq 0.15
\end{equation}
\begin{equation}
	-0.15 \leq u_l \leq 0.15
\end{equation}
\end{subequations}
For all designs considered $T_s = 100$ ms, the number of interior-point iterations $n_{iter} = 15$ and the number of MINRES iterations was set to be equal to corresponding linear system size. Suitability of these parameters was experimentally confirmed based on closed-loop simulation results.

\begin{figure}[tb] 
    \centering
    \begin{minipage}{0.45\textwidth}
       \centering
        \includegraphics[width=\textwidth]{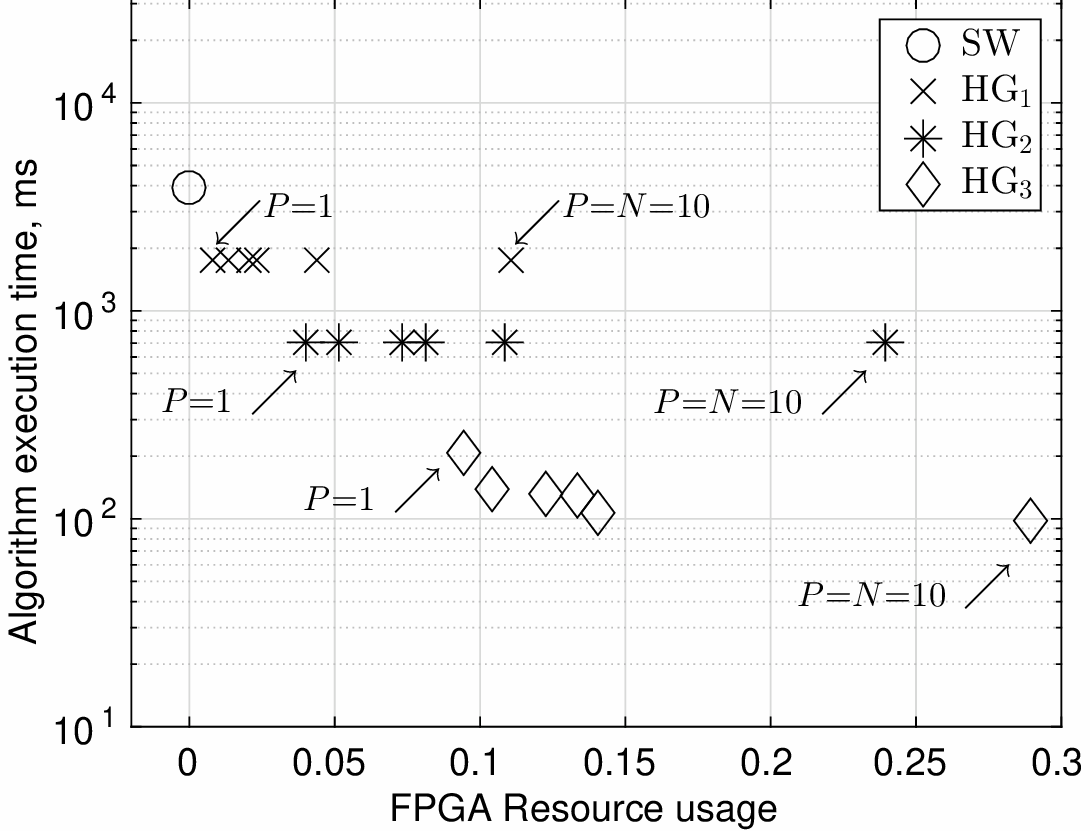} 
        \caption{Trading off algorithm execution time against resource usage by splitting the workload between software and hardware. For all implementations $N = 10$.}
        \label{fig:heterogeneous_tradeoffs}
    \end{minipage}\hfill
    \begin{minipage}{0.45\textwidth}
        \centering
        \includegraphics[width=\textwidth]{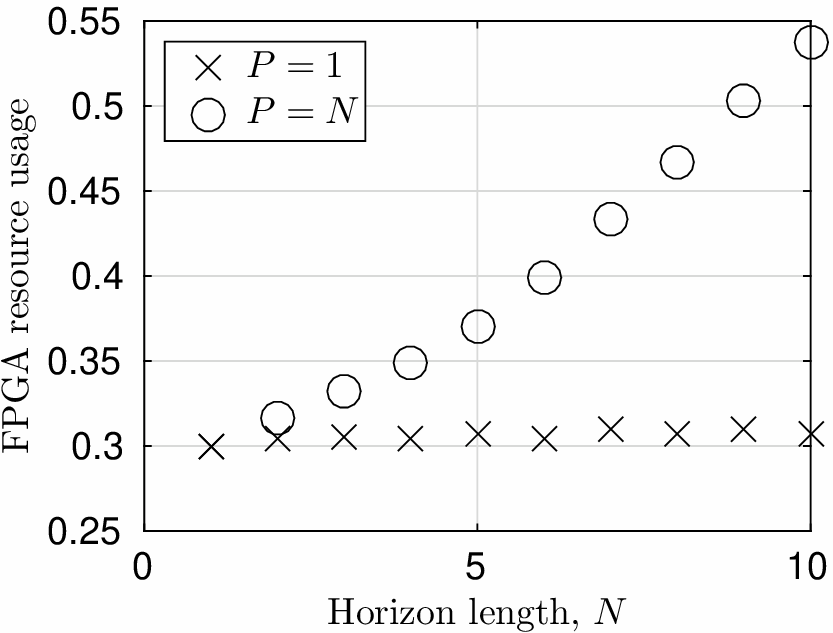} 
        \caption{The impact of horizon length on FPGA resource usage with HG$_3$ implementation.}
        \label{fig:resource_scaling_horizon}
    \end{minipage}
\end{figure}

With the proposed implementation computation time can be traded off against resource usage either by changing the level of parallelism (i.e. the number of MAC units for matrix-vector multiplication) in the hardware accelerator or by shifting MINRES algorithm layers between software or hardware. These tradeoffs are shown in Figure~\ref{fig:heterogeneous_tradeoffs}. 
Resource usage is calculated as a Euclidian norm of a vector that contains the relative utilization of different FPGA resources (LUTs, FFs, DSPs and BRAMs). For each heterogeneous implementation (HG$_1$, HG$_2$ and HG$_3$) the parallelism factor $P$ was varied from 1 to $N$. Note that although $N = 10$, there are only 6 designs for each implementation. This happens because some values of $P$ are identified as inefficient at code generation stage. For this example any design with $P = 6 \textrm{ to } 9$ will lead to the same performance as $P = 5$. We highlight some other conclusions that can be drawn based on Figure~\ref{fig:heterogeneous_tradeoffs}:
\begin{itemize}
	\item Considering execution time and resource usage as two contradicting design objectives within a multi-objective optimization problem, it can be seen that Pareto-optimal designs are achieved by splitting the workload between software and hardware in different ways. This justifies the flexibility of the software-hardware splitting of the proposed implementation.
	\item For HG$_1$ and HG$_2$ implementations increasing matrix-vector multiplication parallelism does not result in any speed up. This happens due to the fact that the Lanczos kernel (implemented on the FPGA) and the remaining part of the MINRES algorithm (implemented on the CPU) run in parallel and the overall execution time is defined by the slowest branch, which is the remaining part of MINRES.
	\item For the test case considered the maximum speedup of hardware-accelerated implementations over a pure software realization is 36x. Since this speedup comes at the price of significantly increased resource usage, it might be reasonable to compare the fastest design against other Pareto-optimal designs before making the final design choice.
\end{itemize}
Increasing the prediction horizon $N$, while improving closed-loop performance, normally results in implementations with high resource consumption. Consider Figure~\ref{fig:resource_scaling_horizon} that visualizes scaling of FPGA resource usage with the prediction horizon for two extreme cases: $P = 1$ and $P = N$. It can be observed that changing the degree of parallelism allows taking control over resource scaling. Note that for sequential processing of KKT matrix blocks ($P = 1$) resource scaling can hardly be noticed. This happens due to the fact that the memory footprint is so small that fixed-size FPGA memory blocks are not fully filled for $N=1$ nor for $N=10$. Further synthesis details can be found in~\ref{sec:synthesis_table}.

\begin{figure}[tb]
        \centering
        \includegraphics[width=1\textwidth]{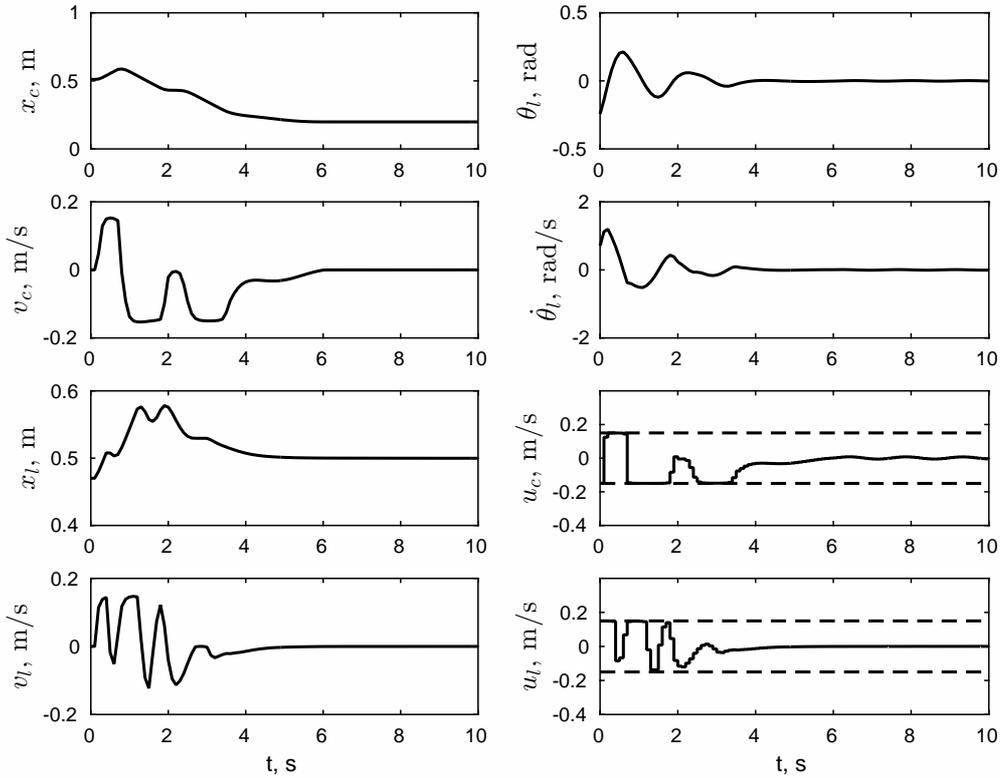} 
        \caption{Processor-in-the-loop simulation results with HG$_3$ implementation. $N=10$, $P=10$.}
        \label{fig:closed_loop_response}
\end{figure}

Finally we present closed-loop simulation results obtained with a HG$_3$ implementation in the loop with Simscape model (Figure~\ref{fig:closed_loop_response}). Starting from the initial condition $ \hat{x} = [ 0.5 \ 0 \ 0.7 \ 0 \ -0.2 \ -0.5 \ -0.15 \ -0.15]^T $ the controller drives the system to the desired point according to the objective~\eqref{eq:objective} while satisfying constraints~\eqref{eq:constraints}.

%% file: conclusion.tex
\section{Conclusions}

This paper presented the following contributions:
\begin{itemize}
	\item A heterogeneous implementation of an  interior-point-based nonlinear predictive controller and experimental validation of the controller in the loop with a gantry crane model.
	\item A new release of Protoip, a software tool for quick prototyping of optimization-based controllers on heterogeneous computers. The tool can be used both for testing the proposed controller and for prototyping new algorithms.
\end{itemize}

The following conclusions can be drawn out of this work:
\begin{itemize}
	\item Accelerating the linear algebra routines in hardware within a heterogeneous computer implementation can result in a significant speedup over a software-only implementation.
	\item Performance of a heterogeneous computer-based implementation can be efficiently traded off against resource usage by shifting the computational workload between the CPU and the FPGA, while varying the amount of parallelism for a given part of an algorithm might be less efficient or even completely gainless.
	\item Offline scheduling for sparse matrix vector multiplication allows avoiding data dependencies and hence building efficient data pipelines, which result in faster implementations.
\end{itemize}

Further work will be focused on automating the design process by formulating the  NMPC design problem as multi-objective optimization problem in order to identify design trade offs in a systematic way as in \citep{bulat_ecc}. The list of possible design variables can include both hardware parameters (e.g.\ parallelization level) and software parameters (e.g.\ horizon length or number of iterations) parameters.

%% file: acknowledgements.tex
\section{Acknowledgements}

This work was funded from the People Programme (Marie Curie Actions) of the European Union's Seventh Framework Programme (FP7/2007-2013) under REA grant agreement no 607957 (TEMPO). The authors also would like to acknowledge the discussions with Mr Junyi Liu, Mr Andrea Zanelli, Dr Rien Quirynen and Dr Milan Vukov, and industrial support from Mathworks.

%% file: appendix_algorithm_details.tex
\section{Remaining details for Algorithm~\ref{alg:ip}} \label{sec:algorith_details}
$$g(\theta_{[k]}) \coloneqq J \theta_{[k]} - d$$
$$\quad  W_{[k]} \coloneqq \Lambda_{[k]}G^{-1}(\theta_{[k]}), $$
where $\Lambda_{[k]}$ and $G(\theta_{[k]})$ are diagonal matrices containing the elements of $\lambda_{[k]}$ and $g(\theta_{[k]})$ respectively.

$$ H_{[k]} \coloneqq \nabla_{\theta} f^{T} (\theta_{[k]}) \nabla_{\theta} f (\theta_{[k]}), \quad \mu = - \sigma \lambda^T_{[k]} g(\theta _{[k]}) $$

$$r_{dual} \coloneqq  \nabla_{\theta} g^{T} (\theta_{[k]}) G^{-1} (\theta_{[k]}) \mu e - \nabla_{\theta} f^{T}(\theta_{[k]}) \nabla_{\theta} f(\theta_{[k]}) - \nabla_{\theta} p^T(\theta_{[k]}) \nu, \quad r_{eq} \coloneqq -p(\theta_{[k]})$$

%% file: appendix_synthesis_results.tex
\section{Synthesis results} \label{sec:synthesis_table}

The synthesis results are presented in Table~\ref{tab:time_comparison}.

\begin{table*}[!h]
\centering
\caption{Algorithm execution times with heterogeneous implementations. For all considered implementations the CPU clock rate is 667\,MHz and the FPGA clock frequency is 166\,MHz. $T_s=100$\,ms, $n_{iter} = 15$, $t$ is the algorithm execution time.}
\rotatebox{90}{
\begin{tabularx}{1.15\textwidth}{l*{18}c}
\toprule
$N$ & $P$ & \lcol{SW} &  \multicolumn{5}{|c}{HG$_1$} & \multicolumn{5}{|c}{HG$_2$} & \multicolumn{5}{|c}{HG$_3$} \\ \hline
 & & \lcol{t, ms} & \lcol{t, ms} & LUT & FF & DSP  & \begin{tabular}[x]{@{}c@{}}\footnotesize{block}\\ \footnotesize{RAM}\end{tabular} & \lcol{t, ms} & LUT & FF & DSP  &\begin{tabular}[x]{@{}c@{}} \footnotesize{block} \\ \footnotesize{RAM} \end{tabular} & \lcol{t, ms} & LUT & FF & DSP  &\begin{tabular}[x]{@{}c@{}} \footnotesize{block} \\ \footnotesize{RAM} \end{tabular} \\ \hline
1 & 1&   72 &  34 & 2242  & 3242  &  8 & 11 & 20 & 6236 & 8429 & 20 & 26 &    9 & 8469 & 11723 & 43 & 33 \\ \hline
2 & 1&  201 &  94 & 2653  & 3726  & 10 & 12 & 49 & 6684 & 8793 & 20 & 26 &   19 & 8868 & 12071 & 43 & 33 \\
2 & 2&   -  &  94 & 2727  & 3859  & 10 & 15 & 49 & 6745 & 8972 & 23 & 33 &   16 & 8960 & 12136 & 43 & 40 \\ \hline
3 & 1&  406 & 180 & 2646  & 3836  & 10 & 11 & 92 & 6502 & 8758 & 20 & 26 &   31 & 8910 & 12084 & 43 & 33 \\
3 & 3&  -   & 180 & 3249  & 4485  & 15 & 19 & 91 & 7080 & 9202 & 26 & 40 &   24 & 9277 & 12264 & 43 & 47 \\ \hline
4 & 1&  654 & 288 & 2637  & 3868  & 10 & 11 &140 & 6542 & 8817 & 20 & 26 &   47 & 8786 & 12143 & 43 & 33 \\
4 & 4&  -   & 290 & 3689  & 5059  & 20 & 25 &141 & 7617 & 9852 & 31 & 47 &   32 & 9636 & 12472 & 43 & 54 \\ \hline
5 & 1&  967 & 421 & 2576  & 3817  & 10 & 12 &202 & 6432 & 8762 & 20 & 27 &   65 & 8976 & 12062 & 43 & 34 \\
5 & 5&  -   & 421 & 4006  & 5714  & 25 & 31 &201 & 8203 &10706 & 36 & 55 &   42 &10037 & 12851 & 43 & 62 \\ \hline
6 & 1& 1350 & 587 & 2547  & 3821  & 10 & 12 &276 & 6439 & 8759 & 20 & 27 &   87 & 8671 & 12064 & 43 & 34 \\
6 & 6&  -   & 594 & 4424  & 6266  & 30 & 36 &276 & 8705 & 1135 & 41 & 62 &   51 &10542 & 13154 & 46 & 69 \\ \hline
7 & 1& 1812 & 781 & 2582  & 3849  & 10 & 13 &362 & 6503 & 8814 & 20 & 28 &  112 & 9064 & 12172 & 43 & 35 \\
7 & 7&  -   & 783 & 4841  & 6931  & 35 & 41 &360 & 9324 &12100 & 46 & 69 &   62 &11077 & 13914 & 51 & 76 \\ \hline
8 & 1& 2323 & 994 & 2564  & 3850  & 10 & 13 &460 & 6499 & 8820 & 20 & 28 &  141 & 8820 & 12172 & 43 & 35 \\
8 & 8&  -   & 998 & 5200  & 7488  & 40 & 46 &459 & 9820 &12759 & 51 & 76 &   73 &11489 & 14573 & 56 & 83 \\ \hline
9 & 1& 2877 &1275 & 2590  & 3863  & 10 & 13 &571 & 6503 & 8825 & 20 & 28 &  172 & 9093 & 12179 & 43 & 35 \\
9 & 9&  -   &1275 & 5651  & 8045  & 45 & 51 &571 &10461 &13401 & 56 & 83 &   84 &12182 & 15222 & 61 & 90 \\ \hline
10& 1& 3523 &1524 & 2616  & 3865  & 10 & 13 &694 & 6499 & 8828 & 20 & 28 &  205 & 8787 & 12174 & 43 & 35 \\
10&10&  -   &1524 & 6005  & 8593  & 50 & 56 &694 &10906 &14044 & 61 & 90 &   96 & 12693& 15864 & 66 & 97 \\ \hline
\bottomrule
\end{tabularx}
}
\label{tab:time_comparison}
\end{table*}

%% file: root.bbl
\begin{thebibliography}{32}
\providecommand{\natexlab}[1]{#1}
\providecommand{\url}[1]{\texttt{#1}}
\expandafter\ifx\csname urlstyle\endcsname\relax
  \providecommand{\doi}[1]{doi: #1}\else
  \providecommand{\doi}{doi: \begingroup \urlstyle{rm}\Url}\fi

\bibitem[Amdahl(1967)]{amdahls_law}
Gene~M. Amdahl.
\newblock Validity of the single processor approach to achieving large scale
  computing capabilities.
\newblock In \emph{Proceedings of the April 18-20, 1967, Spring Joint Computer
  Conference}, AFIPS '67 (Spring), pages 483--485, New York, NY, USA, 1967.
  ACM.
\newblock \doi{10.1145/1465482.1465560}.
\newblock URL \url{http://doi.acm.org/10.1145/1465482.1465560}.

\bibitem[Andersson(2013)]{casadi}
Joel Andersson.
\newblock \emph{{A} {G}eneral-{P}urpose {S}oftware {F}ramework for {D}ynamic
  {O}ptimization}.
\newblock {P}h{D} thesis, Arenberg Doctoral School, KU Leuven, Department of
  Electrical Engineering (ESAT/SCD) and Optimization in Engineering Center,
  Kasteelpark Arenberg 10, 3001-Heverlee, Belgium, October 2013.

\bibitem[Ayala et~al.(2016)Ayala, Sampaio, Mu{\~n}oz, Llanos, Coelho, and
  Jacobi]{fpga_ann_nmpc}
H.~Ayala, R.~Sampaio, D.~M. Mu{\~n}oz, C.~Llanos, L.~Coelho, and R.~Jacobi.
\newblock Nonlinear model predictive control hardware implementation with
  custom-precision floating point operations.
\newblock In \emph{2016 24th Mediterranean Conference on Control and Automation
  (MED)}, pages 135--140, June 2016.
\newblock \doi{10.1109/MED.2016.7535908}.

\bibitem[Betts(2010)]{betts_nlp}
J.T. Betts.
\newblock \emph{Practical Methods for Optimal Control Using Nonlinear
  Programming}.
\newblock Society for Industrial and Applied Mathematics, second edition, 2010.

\bibitem[Boland and Constantinides(2008)]{minres_fpga}
David Boland and George~A Constantinides.
\newblock An {FPGA}-based implementation of the minres algorithm.
\newblock In \emph{2008 International Conference on Field Programmable Logic
  and Applications}, pages 379--384. IEEE, 2008.

\bibitem[Boland and Constantinides(2011)]{boland_memory_bandwidth}
David Boland and George~A. Constantinides.
\newblock Optimizing memory bandwidth use and performance for matrix-vector
  multiplication in iterative methods.
\newblock \emph{ACM Trans. Reconfigurable Technol. Syst.}, 4\penalty0
  (3):\penalty0 22:1--22:14, August 2011.
\newblock ISSN 1936-7406.
\newblock \doi{10.1145/2000832.2000834}.
\newblock URL \url{http://doi.acm.org/10.1145/2000832.2000834}.

\bibitem[Davidon(1991)]{bfgs}
William~C. Davidon.
\newblock Variable metric method for minimization.
\newblock \emph{SIAM Journal on Optimization}, 1\penalty0 (1):\penalty0 1--17,
  1991.
\newblock \doi{10.1137/0801001}.
\newblock URL \url{http://dx.doi.org/10.1137/0801001}.

\bibitem[Debrouwere et~al.(2014)Debrouwere, Vukov, Quirynen, Diehl, and
  Swevers]{acado_crane}
Frederik Debrouwere, Milan Vukov, Rien Quirynen, Moritz Diehl, and Jan Swevers.
\newblock Experimental validation of combined nonlinear optimal control and
  estimation of an overhead crane.
\newblock \emph{{IFAC} Proceedings Volumes}, 47\penalty0 (3):\penalty0 9617 --
  9622, 2014.
\newblock ISSN 1474-6670.
\newblock \doi{10.3182/20140824-6-ZA-1003.01674}.
\newblock 19th {IFAC} World Congress.

\bibitem[Fliess et~al.(1991)Fliess, Levine, and Rouchon]{crane_model}
M.~Fliess, J.~Levine, and P.~Rouchon.
\newblock A simplified approach of crane control via a generalized state-space
  model.
\newblock In \emph{[1991] Proceedings of the 30th IEEE Conference on Decision
  and Control}, pages 736--741 vol.1, Dec 1991.
\newblock \doi{10.1109/CDC.1991.261409}.

\bibitem[Greenbaum(1997)]{greenbaum_iterative_methods}
A.~Greenbaum.
\newblock \emph{Iterative Methods for Solving Linear Systems}.
\newblock Society for Industrial and Applied Mathematics, 1997.
\newblock \doi{10.1137/1.9781611970937}.
\newblock URL \url{http://epubs.siam.org/doi/abs/10.1137/1.9781611970937}.

\bibitem[Hartley et~al.(2014)Hartley, Jerez, Suardi, Maciejowski, Kerrigan, and
  Constantinides]{hartley_fpga}
E.N. Hartley, J.L. Jerez, A.~Suardi, Jan~M. Maciejowski, E.C. Kerrigan, and
  G.~Constantinides.
\newblock {Predictive Control using an FPGA with Application to Aircraft
  Control}.
\newblock \emph{IEEE Transactions on Control Systems Technology}, 22\penalty0
  (3):\penalty0 1006--1017, May 2014.

\bibitem[Hindmarsh et~al.(2005)Hindmarsh, Brown, Grant, Lee, Serban, Shumaker,
  and Woodward]{sundials}
Alan~C Hindmarsh, Peter~N Brown, Keith~E Grant, Steven~L Lee, Radu Serban,
  Dan~E Shumaker, and Carol~S Woodward.
\newblock {SUNDIALS}: Suite of nonlinear and differential/algebraic equation
  solvers.
\newblock \emph{ACM Transactions on Mathematical Software (TOMS)}, 31\penalty0
  (3):\penalty0 363--396, 2005.

\bibitem[Houska et~al.(2011)Houska, Ferreau, and Diehl]{acado_reference}
B.~Houska, H.J. Ferreau, and M.~Diehl.
\newblock {An Auto-Generated Real-Time Iteration Algorithm for Nonlinear {MPC}
  in the Microsecond Range}.
\newblock \emph{Automatica}, 47\penalty0 (10):\penalty0 2279--2285, 2011.
\newblock \doi{10.1016/j.automatica.2011.08.020}.

\bibitem[Inc(2012)]{zedboard}
Inc.
\newblock {ZedBoard} {(Zynq} evaluation and development) hardware user's guide,
  September 2012.

\bibitem[Janka et~al.(2016)Janka, Kirches, Sager, and W{\"a}chter]{block_bfgs}
Dennis Janka, Christian Kirches, Sebastian Sager, and Andreas W{\"a}chter.
\newblock An sr1/bfgs sqp algorithm for nonconvex nonlinear programs with
  block-diagonal hessian matrix.
\newblock \emph{Mathematical Programming Computation}, 8\penalty0 (4):\penalty0
  435--459, 2016.
\newblock URL \url{http://dx.doi.org/10.1007/s12532-016-0101-2}.

\bibitem[Jerez et~al.(2012)Jerez, Ling, Constantinides, and
  Kerrigan]{jerezetalIET:2012}
J.~L. Jerez, K.~V. Ling, G.~A. Constantinides, and E.~C. Kerrigan.
\newblock Model predictive control for deeply pipelined field-programmable gate
  array implementation: algorithms and circuitry.
\newblock \emph{IET Control Theory \& Applications}, 6:\penalty0
  1029--1041(12), May 2012.
\newblock ISSN 1751-8644.

\bibitem[Jerez et~al.(2015)Jerez, Constantinides, and
  Kerrigan]{lanczos_scaling}
J.~L. Jerez, G.~A. Constantinides, and E.~C. Kerrigan.
\newblock A low complexity scaling method for the lanczos kernel in fixed-point
  arithmetic.
\newblock \emph{IEEE Transactions on Computers}, 64\penalty0 (2):\penalty0
  303--315, Feb 2015.
\newblock ISSN 0018-9340.
\newblock \doi{10.1109/TC.2013.162}.

\bibitem[Khan et~al.(2011)Khan, Hafeez, Mirza, and u.~Ain]{fpga_pci}
F.~A. Khan, Z.~Hafeez, A.~Mirza, and Q.~u.~Ain.
\newblock Design of fpga based daq card using pci express protocol.
\newblock In \emph{2011 IEEE 14th International Multitopic Conference}, pages
  211--216, Dec 2011.
\newblock \doi{10.1109/INMIC.2011.6151475}.

\bibitem[Khusainov et~al.()Khusainov, Kerrigan, Suardi, and
  Constantinides]{bulat_heterogeneous_nmpc}
B.~Khusainov, E.C. Kerrigan, A.~Suardi, and G.A. Constantinides.
\newblock Nonlinear predictive control on a heterogeneous computing platform.
\newblock IFAC World Congress 2017.
\newblock URL \url{http://hdl.handle.net/10044/1/45094}.

\bibitem[Khusainov et~al.(2016)Khusainov, Kerrigan, and
  Constantinides]{bulat_ecc}
B~Khusainov, EC~Kerrigan, and GA~Constantinides.
\newblock Multi-objective co-design for model predictive control with an
  {FPGA}.
\newblock In \emph{European Control Conference 2016}. IEEE, 2016.
\newblock URL \url{http://hdl.handle.net/10044/1/30637}.

\bibitem[Lofberg(2004)]{yalmip}
J.~Lofberg.
\newblock Yalmip : a toolbox for modeling and optimization in matlab.
\newblock In \emph{2004 IEEE International Conference on Robotics and
  Automation (IEEE Cat. No.04CH37508)}, pages 284--289, Sept 2004.
\newblock \doi{10.1109/CACSD.2004.1393890}.

\bibitem[Mat(2015)]{matlab_bib}
\emph{MATLAB version 9.2 (R2017a)}.
\newblock The Mathworks, Inc., Natick, Massachusetts, 2015.

\bibitem[Nocedal et~al.(2005)Nocedal, W\"achter, and Waltz]{barrier_ip}
J.~Nocedal, A.~W\"achter, and R.~A. Waltz.
\newblock Adaptive barrier strategies for nonlinear interior methods.
\newblock Technical Report RC 23563, IBM Watson Research Center, Yorktown
  Heights, NY, USA, 2005.

\bibitem[Paige and Saunders(1975)]{minres}
C.~C. Paige and M.~A. Saunders.
\newblock Solution of sparse indefinite systems of linear equations.
\newblock \emph{SIAM Journal on Numerical Analysis}, 12\penalty0 (4):\penalty0
  617--629, 1975.
\newblock \doi{10.1137/0712047}.
\newblock URL \url{http://dx.doi.org/10.1137/0712047}.

\bibitem[Peyrl et~al.(2015)Peyrl, Ferreau, and Kouzoupis]{peyrl_hybrid_nmpc}
Helfried Peyrl, Hans~Joachim Ferreau, and Dimitris Kouzoupis.
\newblock A hybrid hardware implementation for nonlinear model predictive
  control.
\newblock \emph{IFAC-PapersOnLine}, 48\penalty0 (23):\penalty0 87 -- 93, 2015.
\newblock ISSN 2405-8963.
\newblock \doi{http://dx.doi.org/10.1016/j.ifacol.2015.11.266}.

\bibitem[Rawlings and Mayne(2009)]{mayne_book}
J.B. Rawlings and D.Q. Mayne.
\newblock \emph{Model Predictive Control: Theory and Design}.
\newblock Nob Hill Pub., 2009.
\newblock ISBN 9780975937709.

\bibitem[Saad and Schultz(1986)]{gmres}
Youcef Saad and Martin~H. Schultz.
\newblock Gmres: A generalized minimal residual algorithm for solving
  nonsymmetric linear systems.
\newblock \emph{SIAM Journal on Scientific and Statistical Computing},
  7\penalty0 (3):\penalty0 856--869, 1986.
\newblock \doi{10.1137/0907058}.
\newblock URL \url{http://dx.doi.org/10.1137/0907058}.

\bibitem[Shahzad et~al.(2012)Shahzad, Kerrigan, and
  Constantinides]{shahzadetal:2012}
Amir Shahzad, Eric~C. Kerrigan, and George~A. Constantinides.
\newblock A stable and efficient method for solving a convex quadratic program
  with application to optimal control.
\newblock \emph{SIAM Journal on Optimization}, 22\penalty0 (4):\penalty0
  1369--1393, 2012.
\newblock \doi{10.1137/11082960X}.
\newblock URL \url{http://dx.doi.org/10.1137/11082960X}.

\bibitem[Shukla et~al.()Shukla, Khusainov, Kerrigan, and Jones]{protosplit}
H.~Shukla, B.~Khusainov, E.C. Kerrigan, and C.N. Jones.
\newblock Software and hardware code generation for predictive control using
  splitting methods.
\newblock IFAC World Congress 2017.
\newblock URL \url{http://hdl.handle.net/10044/1/45093}.

\bibitem[Suardi et~al.(2015)Suardi, Constantinides, and Kerrigan]{protoip}
A.~Suardi, G.~A. Constantinides, and E.~C. Kerrigan.
\newblock Software development kit for {FPGA}: A fast {FPGA} prototyping tool
  for embedded optimization.
\newblock In \emph{European Control Conference}, 2015.

\bibitem[W{\"a}chter and Biegler(2006)]{ipopt}
Andreas W{\"a}chter and Lorenz~T. Biegler.
\newblock On the implementation of an interior-point filter line-search
  algorithm for large-scale nonlinear programming.
\newblock \emph{Mathematical Programming}, 106\penalty0 (1):\penalty0 25--57,
  2006.
\newblock ISSN 1436-4646.
\newblock \doi{10.1007/s10107-004-0559-y}.
\newblock URL \url{http://dx.doi.org/10.1007/s10107-004-0559-y}.

\bibitem[Xu et~al.(2016)Xu, Chen, Gong, and Mei]{fpga_pso_nmpc}
F.~Xu, H.~Chen, X.~Gong, and Q.~Mei.
\newblock Fast nonlinear model predictive control on {FPGA} using particle
  swarm optimization.
\newblock \emph{IEEE Transactions on Industrial Electronics}, 63\penalty0
  (1):\penalty0 310--321, Jan 2016.
\newblock ISSN 0278-0046.
\newblock \doi{10.1109/TIE.2015.2464171}.

\end{thebibliography}
